\documentclass{article}

\usepackage{arxiv}

\usepackage[utf8]{inputenc}
\usepackage[T1]{fontenc}
\usepackage{hyperref}
\usepackage{url}
\usepackage{booktabs}
\usepackage{amsfonts}
\usepackage{amssymb}
\usepackage{nicefrac}
\usepackage{microtype}
\usepackage{graphicx}
\usepackage[numbers,sort&compress]{natbib}
\usepackage{xcolor}
\usepackage{listings}
\usepackage{subcaption}
\usepackage{multirow}
\usepackage{tikz}
\usepackage{pgfplots}

\pgfplotsset{compat=1.18} 
\usepgfplotslibrary{fillbetween}
\usetikzlibrary{arrows.meta, positioning, fit, backgrounds, calc, decorations.pathreplacing}
\usepackage[most]{tcolorbox}
\usepackage{fontawesome5}
\tcbuselibrary{listings, breakable}
\usepackage{cleveref}
\usepackage{orcidlink}

\definecolor{colNeonPink}{HTML}{FF2BD6}
\definecolor{colHotMagenta}{HTML}{FF008C}
\definecolor{colElectricBlue}{HTML}{00E5FF}
\definecolor{colAcidPurple}{HTML}{9D4EDD}
\definecolor{colDeepBlack}{HTML}{0A0A0F}
\definecolor{colSingle}{HTML}{FF2BD6}    % Neon Pink — single agent, L0 spec
\definecolor{colSplit}{HTML}{00E5FF}     % Electric Blue — split agent, L3 spec
\definecolor{colRecall}{HTML}{FF008C}    % Hot Magenta — recall, resolve
\definecolor{colPrecision}{HTML}{9D4EDD} % Acid Purple — precision
\definecolor{colGap}{HTML}{9D4EDD}       % Acid Purple — gap annotations
\definecolor{colType}{HTML}{FF2BD6}      % Neon Pink — type conflicts
\definecolor{colState}{HTML}{00E5FF}     % Electric Blue — state conflicts
\definecolor{colProtocol}{HTML}{9D4EDD}  % Acid Purple — protocol conflicts
\definecolor{colReturn}{HTML}{FF008C}    % Hot Magenta — return conflicts

\newtcolorbox{promptbox}[1][]{%
  enhanced, breakable,
  colback=colAcidPurple!6, colframe=colAcidPurple!45,
  fontupper=\ttfamily\scriptsize,
  arc=2pt, boxrule=0.6pt,
  left=6pt, right=6pt, top=4pt, bottom=4pt,
  title={\sffamily\small\bfseries #1},
  coltitle=colNeonPink!85!black, fonttitle=\sffamily\small\bfseries,
  attach boxed title to top left={yshift=-2mm, xshift=4mm},
  boxed title style={colback=white, colframe=colAcidPurple!45,
                     arc=1pt, boxrule=0.4pt},
}

\newtcolorbox{codebox}[2][]{%
  enhanced, breakable,
  colback=black!3, colframe=#2!50!black,
  fontupper=\ttfamily\scriptsize,
  arc=2pt, boxrule=0.5pt,
  left=6pt, right=6pt, top=4pt, bottom=4pt,
  title={\sffamily\small\bfseries #1},
  coltitle=#2!80!black, fonttitle=\sffamily\small\bfseries,
  attach boxed title to top left={yshift=-2mm, xshift=4mm},
  boxed title style={colback=white, colframe=#2!50!black,
                     arc=1pt, boxrule=0.4pt},
}

\newtcolorbox{conflictbox}[1][]{%
  enhanced, breakable,
  colback=colHotMagenta!4, colframe=colHotMagenta!60,
  fontupper=\ttfamily\scriptsize,
  arc=2pt, boxrule=0.6pt,
  left=6pt, right=6pt, top=4pt, bottom=4pt,
  title={\sffamily\small\bfseries #1},
  coltitle=colHotMagenta!90!black, fonttitle=\sffamily\small\bfseries,
  attach boxed title to top left={yshift=-2mm, xshift=4mm},
  boxed title style={colback=white, colframe=colHotMagenta!60,
                     arc=1pt, boxrule=0.4pt},
}

\newtcolorbox{specbox}[2][]{%
  enhanced, breakable,
  colback=#2!3, colframe=#2!45!black,
  fontupper=\ttfamily\scriptsize,
  arc=2pt, boxrule=0.5pt,
  left=6pt, right=6pt, top=4pt, bottom=4pt,
  title={\sffamily\small\bfseries #1},
  coltitle=#2!70!black, fonttitle=\sffamily\small\bfseries,
  attach boxed title to top left={yshift=-2mm, xshift=4mm},
  boxed title style={colback=white, colframe=#2!45!black,
                     arc=1pt, boxrule=0.4pt},
}

\newcommand{\eg}{\textit{e.g.}}

\newcommand{\speclevel}[1]{\textsc{L#1}}

\renewcommand{\shorttitle}{The Specification Gap}

\title{The Specification Gap: Coordination Failure Under Partial Knowledge in Code Agents}

\date{}

\author{
  Camilo Chac\'on Sartori\orcidlink{0000-0002-8543-9893}\thanks{Corresponding author.}\\
  Catalan Institute of Nanoscience and Nanotechnology (ICN2), CSIC and BIST,\\
  Campus UAB, Bellaterra, Barcelona, Spain\\
  \texttt{camilo.chacon@icn2.cat}\\[6pt]
  \faGithub\ \url{https://github.com/camilochs/the_specification_gap}
}

\hypersetup{
  pdftitle={The Specification Gap: Coordination Failure Under Partial Knowledge in Code Agents},
  pdfsubject={cs.SE, cs.AI},
  pdfauthor={Camilo Chacon Sartori},
  pdfkeywords={multi-agent systems, code generation, specification, coordination, LLM},
}

\begin{document}
\maketitle

% ============================================================
\begin{abstract}
When multiple LLM-based code agents independently implement parts of the same class, they must agree on shared internal representations---even when the specification leaves those choices implicit. We study this coordination problem across 51 class-generation tasks, progressively stripping specification detail from full docstrings (\speclevel{0}) to bare signatures (\speclevel{3}), and introducing opposing structural biases (lists vs.\ dictionaries) to stress-test integration. Three findings emerge. First, a persistent \emph{specification gap}: two-agent integration accuracy drops from 58\% to 25\% as detail is removed, while a single-agent baseline degrades more gracefully (89\% to 56\%), leaving a 25--39pp coordination gap that is consistent across two Claude models (Sonnet, Haiku) and three independent runs. Second, an AST-based conflict detector achieves 97\% precision at the weakest specification level without additional LLM calls, yet a factorial recovery experiment shows that restoring the full specification alone recovers the single-agent ceiling (89\%), while providing conflict reports adds no measurable benefit. Third, decomposing the gap into coordination cost (+16pp) and information asymmetry (+11pp) suggests that the two effects are independent and approximately additive---the gap is not merely a consequence of hidden information but reflects the difficulty of producing compatible code without shared decisions. These results support a \emph{specification-first} view of multi-agent code generation: richer specifications are both the primary coordination mechanism and the sufficient recovery instrument.
\end{abstract}


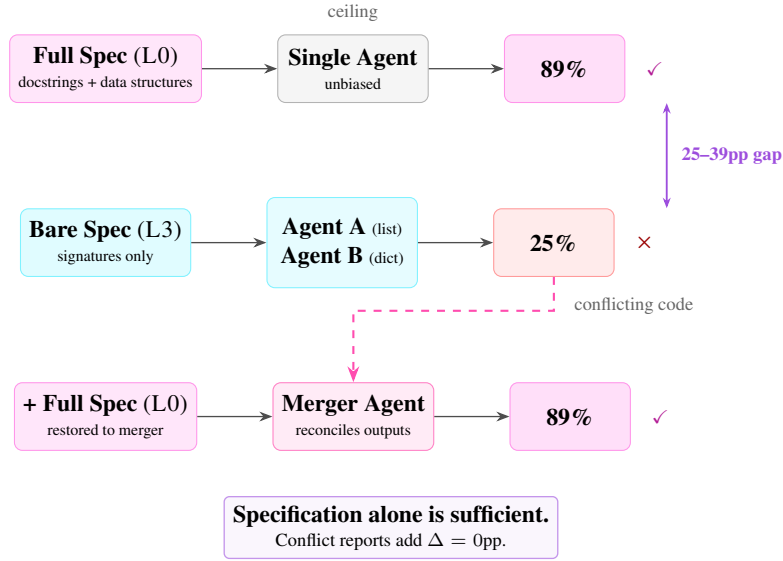
\begin{figure*}[t]
  \centering
  \begin{tikzpicture}[
    box/.style={draw, rounded corners=3pt, minimum height=0.9cm,
                font=\small, align=center, line width=0.4pt},
    specbox/.style={box, minimum width=2.2cm},
    agentbox/.style={box, minimum width=2.0cm},
    resultbox/.style={box, minimum width=1.6cm, font=\small\bfseries},
    arr/.style={-{Stealth[length=5pt]}, line width=0.5pt, black!70},
    bigarr/.style={-{Stealth[length=6pt]}, line width=0.8pt},
    lbl/.style={font=\scriptsize, text=black!60},
  ]
    % ── Row 1: Single agent baseline ──
    \node[specbox, fill=colSingle!12, draw=colSingle!50]
      (spec0) {\textbf{Full Spec} (\speclevel{0})\\[-1pt]{\tiny docstrings + data structures}};
    \node[agentbox, fill=black!4, draw=black!30, right=1.0cm of spec0]
      (single) {\textbf{Single Agent}\\[-1pt]{\tiny unbiased}};
    \node[resultbox, fill=colSingle!15, draw=colSingle!50, right=1.0cm of single]
      (pass1) {89\%};
    \node[font=\scriptsize, colSingle!70!black, right=0.15cm of pass1] {\checkmark};

    \draw[arr] (spec0) -- (single);
    \draw[arr] (single) -- (pass1);
    \node[lbl, above=0.05cm of single] {ceiling};

    % ── Row 2: Split agents — the problem ──
    \node[specbox, fill=colSplit!12, draw=colSplit!50, below=1.4cm of spec0]
      (spec3) {\textbf{Bare Spec} (\speclevel{3})\\[-1pt]{\tiny signatures only}};
    \node[agentbox, fill=colSplit!10, draw=colSplit!50,
          right=1.0cm of spec3, minimum height=1.2cm]
      (split) {\textbf{Agent A} {\tiny(list)}\\[1pt]\textbf{Agent B} {\tiny(dict)}};
    \node[resultbox, fill=red!8, draw=red!40, right=1.0cm of split]
      (fail) {25\%};
    \node[font=\scriptsize, red!60!black, right=0.15cm of fail] {$\boldsymbol{\times}$};

    \draw[arr] (spec3) -- (split);
    \draw[arr] (split) -- (fail);

    % ── The gap annotation (right of results) ──
    \path ([xshift=0.55cm]pass1.east) coordinate (gapX);
    \draw[{Stealth[length=4pt]}-{Stealth[length=4pt]}, colGap, line width=0.7pt]
      (gapX |- pass1.south) -- (gapX |- fail.north)
      node[midway, right=2pt, font=\scriptsize\bfseries, colGap] {25--39pp gap};

    % ── Row 3: Recovery ──
    \node[specbox, fill=colSingle!12, draw=colSingle!50, below=1.4cm of spec3]
      (specR) {\textbf{+ Full Spec} (\speclevel{0})\\[-1pt]{\tiny restored to merger}};
    \node[agentbox, fill=colRecall!8, draw=colRecall!50, right=1.0cm of specR]
      (merger) {\textbf{Merger Agent}\\[-1pt]{\tiny reconciles outputs}};
    \node[resultbox, fill=colSingle!15, draw=colSingle!50, right=1.0cm of merger]
      (recover) {89\%};
    \node[font=\scriptsize, colSingle!70!black, right=0.15cm of recover] {\checkmark};

    \draw[arr] (specR) -- (merger);
    \draw[arr] (merger) -- (recover);

    % Arrow: failed outputs feed into merger (route right to avoid overlap)
    \draw[bigarr, colRecall!70, dashed]
      (fail.south) -- ++(0, -0.45cm) -| (merger.north);
    \node[lbl, anchor=north west] at ([xshift=0.15cm, yshift=-0.15cm]fail.south)
      {conflicting code};

    % ── Key insight box ──
    \node[draw=colGap!60, fill=colGap!5, rounded corners=2pt,
          font=\small, align=center, line width=0.5pt,
          below=0.6cm of merger, xshift=0.5cm]
      (insight) {\textbf{Specification alone is sufficient.}\\[-1pt]
                 {\scriptsize Conflict reports add $\Delta = 0$pp.}};

  \end{tikzpicture}
  \caption{The specification gap and its resolution. \textbf{Top:} A single agent with the full specification achieves 89\% pass rate. \textbf{Middle:} Two biased agents with bare specifications produce conflicting code (25\%). \textbf{Bottom:} Restoring the full specification to a merger agent recovers 89\%---matching the single-agent ceiling---while conflict reports add nothing. The specification is both the cause of failure and the sufficient instrument of recovery.}
  \label{fig:graphical-abstract}
\end{figure*}

% ============================================================
\section{Introduction}

Large language models (LLMs) have demonstrated strong performance on single-agent code generation benchmarks~\citep{chen2021codex, li2023starcoder, roziere2024codellama}. A natural next step is to move from one model to \emph{multi-agent} systems, where several LLM-based agents collaborate to produce larger or more complex software artifacts~\citep{hong2024metagpt, qian2024chatdev}. In this paper, we focus on \emph{code agents}: LLM agents specialized for code generation that receive a specification and produce source code. The promise is straightforward: divide a class across agents, let each implement a subset of methods, and then merge the outputs. For empirical software engineering, this raises a concrete question about coordination under decomposition: when independently generated pieces of a software artifact must later be integrated, what information is necessary for them to remain compatible?

But division of labor introduces a coordination problem. When two code agents independently implement methods of the same class, they must agree on shared internal state---the data structures initialized in the class constructor (\texttt{\_\_init\_\_} in Python)---without communicating at runtime. If Agent~A stores records as a list of tuples and Agent~B expects a dictionary keyed by ID, their individually correct methods will fail when integrated. This is the problem of \emph{partial knowledge}: each code agent sees only part of the specification and must infer the rest. Importantly, the errors that arise from this coordination failure are architecturally distinct from those produced by individual agents. Recent philosophical analysis has argued that AI code generation systems exhibit error profiles rooted in their stochastic architecture rather than in human-cognitive limitations~\citep{Sartori2026}; multi-agent decomposition introduces a further error dimension---structural incompatibility between independently generated outputs---that neither single-agent systems nor human teams typically face in the same form.

In traditional software engineering, this problem is solved by \emph{specifications}. Parnas's information-hiding principle~\citep{parnas1972criteria} and Meyer's Design by Contract~\citep{meyer1992dbc} both argue that module interfaces must be precise enough to constrain implementations. But how precise is ``precise enough'' for code agents?

We answer this question empirically. We construct a controlled experiment with one independent variable---\emph{specification completeness}---and measure its effect on multi-agent integration success. Our specification levels range from \speclevel{0} (full docstrings with doctests and explicit data-structure references) to \speclevel{3} (bare method signatures only). We assign two code agents \emph{opposing} implementation biases: one prefers lists, the other dictionaries. A single unbiased agent provides a ceiling at each level. The goal is not to compare specific proprietary models as a benchmark; it is to identify whether a coordination failure mode appears systematically once shared specifications become incomplete.

\Cref{fig:graphical-abstract} summarizes the central phenomenon studied in the paper: specification loss induces a large integration gap, and restoring the full specification is sufficient to recover from it.

Our key findings are:

\begin{enumerate}
  \item \textbf{Specifications are coordination mechanisms.} Multi-agent integration success degrades monotonically from 58.2\% (\speclevel{0}) to 24.6\% (\speclevel{3}) as specification detail is removed. Explicit data-structure references in docstrings override agent biases.

  \item \textbf{The coordination tax is persistent.} The gap between single-agent and multi-agent performance remains significant at every specification level (25--39pp across two models), never narrowing to zero. Better specifications reduce the coordination cost but do not eliminate it in our setting---a substantial penalty persists even with full docstrings.

  \item \textbf{Conflicts are detectable at zero cost.} An AST-based detector identifies type and state conflicts between agent outputs before integration, with precision improving from 43.5\% at \speclevel{0} to 96.7\% at \speclevel{3}---becoming most reliable where specifications are weakest and conflicts most structural.

  \item \textbf{Specification is the sufficient recovery instrument.} A $2\times 2$ factorial experiment ($n=53$ tasks) reveals that providing the full specification to a merger agent restores 88.9\% pass rate---matching the single-agent ceiling (88.3\%)---while conflict reports alone add $\Delta = 0.0$pp and slightly hurt ($-6.6$pp) when the full specification is already available.
\end{enumerate}

Together, these findings connect classical software engineering theory to modern AI practice. They suggest that code agents need specification-aware orchestration---not just task decomposition or post-hoc conflict detection---to coordinate effectively under partial knowledge. More broadly, they frame multi-agent code generation as an empirical software engineering problem about interfaces, integration, and coordination contracts rather than only as a model capability benchmark.

Beyond these empirical findings, we contribute a reusable experimental methodology for studying multi-agent coordination. Our design combines five elements: (i)~a controlled benchmark with nested specification levels that isolate a single variable; (ii)~a specification-ablation protocol that degrades information monotonically; (iii)~a zero-cost AST conflict detector for pre-integration diagnostics; (iv)~a factorial recovery experiment that separates specification quality from conflict information; and (v)~a $2{\times}2$ init-visibility decomposition that disentangles coordination cost from information asymmetry. Each element can be adapted to other multi-agent settings, languages, or task types.

The paper unfolds as follows. \Cref{sec:related} reviews prior work on multi-agent code generation, specification theory, and conflict detection. \Cref{sec:design} details the experimental design, including the four specification levels, the agent configuration, and the $2{\times}2$ factorial recovery experiment. \Cref{sec:results} reports the results for all four research questions. \Cref{sec:discussion} examines the implications of the findings, cross-model replication, and threats to validity. Finally, \Cref{sec:conclusion} concludes the paper with practical recommendations for code agent systems.

% ============================================================
\section{Background and Related Work}
\label{sec:related}

\subsection{Multi-Agent Code Generation}

Recent systems divide code generation across multiple LLM agents. MetaGPT~\citep{hong2024metagpt} assigns software engineering roles (architect, engineer, tester) to agents communicating through structured artifacts. ChatDev~\citep{qian2024chatdev} simulates a software company with chat-based agent interaction. CodeAgent~\citep{tang2024codeagent} integrates external tools for repository-level tasks. More recent work extends this line with adaptive planning for function-level generation~\citep{zhu2025adacoder}, simulation-driven planning and debugging loops~\citep{islam2025codesim}, and decentralized collaboration over shared Git artifacts~\citep{huang2025evogit}.

These systems demonstrate the potential of multi-agent code generation, but they do not isolate how specification quality shapes coordination. Our work complements them with a controlled experiment that manipulates specification completeness as a single independent variable.

\subsection{Specification and Coordination in Software Engineering}

Parnas~\citep{parnas1972criteria} established that modular decomposition requires precise interface specifications to enable independent development. Meyer's Design by Contract~\citep{meyer1992dbc} formalized this idea through preconditions, postconditions, and invariants. Together, these classical principles frame specifications as coordination contracts between independently developed modules.

Conway's Law~\citep{conway1968committees} predicts that system structure mirrors organizational communication structure. In multi-agent LLM systems, the ``organization'' is the set of agents, and the ``communication'' is the shared specification. Our experiment tests whether specification quality predicts the degree to which independent agents produce compatible code.

\subsection{LLM Code Generation Benchmarks}

HumanEval~\citep{chen2021codex} and MBPP~\citep{austin2021program} evaluate single-function generation. ClassEval~\citep{du2024classeval} extends this to class-level generation with inter-method dependencies, making it suitable for studying multi-agent coordination. We build our benchmark (AmbigClass) on ClassEval tasks, generating specification variants at four completeness levels.

\subsection{Conflict Detection in Software}

Static analysis for type conflicts~\citep{engler2001bugs}, merge-conflict detection~\citep{mens2002state}, and program analysis for integration errors~\citep{binkley1995program} are well-studied in traditional software. We adapt these ideas to LLM-generated code, using AST-based analysis to detect type mismatches and state-access conflicts between agent outputs before integration.

% ============================================================
\section{Experimental Design}
\label{sec:design}

Our experiment isolates the effect of specification completeness on multi-agent code integration. We assign two biased LLM agents disjoint subsets of methods from the same class, vary the amount of structural information each agent receives, and measure whether their independently generated code can be merged into a passing implementation. A single-agent baseline, which always sees the full class body, provides the performance ceiling. This design lets us separate specification-driven coordination failures from intrinsic task difficulty, and test whether post-hoc conflict detection or richer specifications better enable recovery.

\subsection{Research Questions}

Four research questions progressively move from characterizing the specification gap to understanding its causes and potential remedies:

\begin{description}
  \item[RQ1] Does specification completeness predict multi-agent integration success?
  \item[RQ2] How does the multi-agent coordination penalty vary across specification levels?
  \item[RQ3] Can AST-based conflict detection provide an observable signal of specification inadequacy?
  \item[RQ4] What enables recovery from multi-agent integration failures: specification quality, conflict information, or their combination?
\end{description}

\subsection{Independent Variable: Specification Level}

We define four nested specification levels, each providing strictly less information than the previous (\Cref{tab:spec-levels}):

\begin{table}[h]
  \caption{Information content at each specification level.}
  \label{tab:spec-levels}
  \centering
  \begin{tabular}{lcccc}
    \toprule
    Information             & \speclevel{0} & \speclevel{1} & \speclevel{2} & \speclevel{3} \\
    \midrule
    Method signatures       & \checkmark & \checkmark & \checkmark & \checkmark \\
    Full docstrings         & \checkmark & \checkmark & simplified & --- \\
    Doctest examples        & \checkmark & ---        & ---        & --- \\
    Edge-case behavior      & \checkmark & \checkmark & ---        & --- \\
    Data-structure references & \checkmark & \checkmark & ---      & --- \\
    \bottomrule
  \end{tabular}
\end{table}

The critical transition is between \speclevel{1} and \speclevel{2}: \speclevel{0}/\speclevel{1} contain explicit references to data structures (\eg, ``add to the \texttt{job\_listings} list''), while \speclevel{2}/\speclevel{3} use abstract language (``Publish positions'').

\subsection{Conditions}

For each task $t$ and specification level $\ell$, we evaluate three conditions:

\begin{description}
  \item[Single($\ell$)] One unbiased agent receives the skeleton at level $\ell$ \emph{with} the \texttt{\_\_init\_\_} body intact. This provides the ceiling: the agent sees the data structures in code.

  \item[Split($\ell$)] Two biased agents each receive the skeleton at level $\ell$ \emph{without} the \texttt{\_\_init\_\_} body (replaced with \texttt{pass}). Agent~A is instructed to prefer lists; Agent~B prefers dictionaries. Each implements a disjoint subset of methods. Their outputs are merged via textual concatenation of method bodies.

  \item[Conflicts($\ell$)] The AST-based conflict detector analyzes the two agent outputs before integration, reporting type, state, and protocol conflicts.
\end{description}

The key design choice: \emph{stripping \texttt{\_\_init\_\_} for split agents only}. This forces split agents to infer data structures from the specification (or default to their bias), while single agents always know the ground truth.

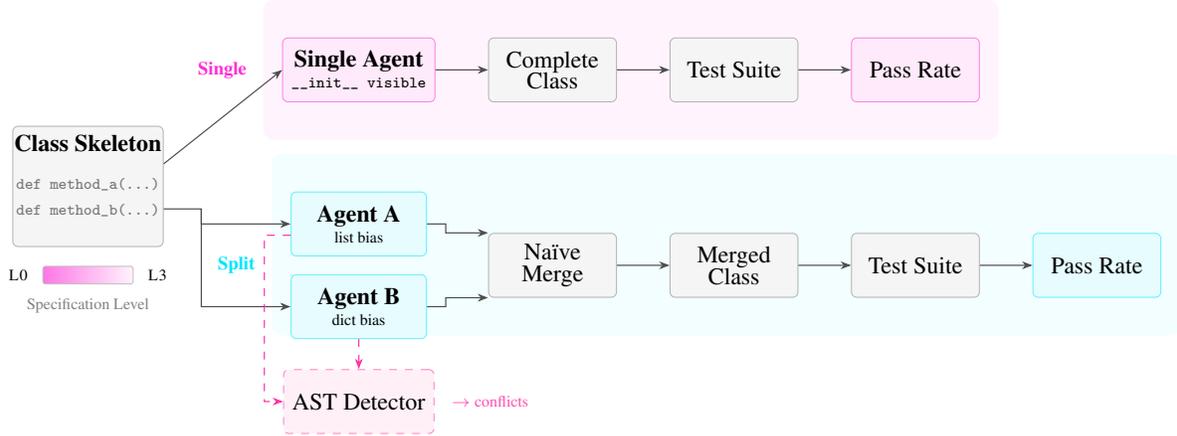
\begin{figure*}[t]
  \centering
  \begin{tikzpicture}[
    box/.style={draw, rounded corners=2pt, minimum height=0.85cm,
                minimum width=1.7cm, font=\small, align=center,
                line width=0.35pt},
    bluebox/.style={box, fill=colSingle!10, draw=colSingle!60},
    orangebox/.style={box, fill=colSplit!10, draw=colSplit!60},
    graybox/.style={box, fill=black!4, draw=black!30},
    arr/.style={-{Stealth[length=4pt]}, line width=0.45pt, black!70},
    darr/.style={-{Stealth[length=4pt]}, line width=0.45pt, dashed, colRecall!80},
    lane/.style={rounded corners=4pt, line width=0pt},
  ]
    % ── Coordinates for lanes ──
    \coordinate (origin) at (0,0);

    % ── Skeleton input (left) ──
    \node[graybox, minimum height=1.6cm, minimum width=2.0cm] at (0,0) (skel) {};
    \node[font=\footnotesize\bfseries, anchor=north, inner sep=0pt]
          at ([yshift=-3pt]skel.north) {Class Skeleton};
    \node[font=\tiny\ttfamily, text=black!55] at ([yshift=1pt]skel.center)
         {def method\_a(...)};
    \node[font=\tiny\ttfamily, text=black!55] at ([yshift=-9pt]skel.center)
         {def method\_b(...)};

    % Spec level badge (below skeleton, centered)
    \node[below=0.25cm of skel] (specbar-center) {};
    \node[minimum width=1.2cm, minimum height=0.22cm,
          left color=colSingle!65, right color=colSingle!6,
          draw=black!25, line width=0.25pt, rounded corners=1pt]
          at (specbar-center) (specbar) {};
    \node[font=\tiny, left=2pt of specbar.west, anchor=east] {\speclevel{0}};
    \node[font=\tiny, right=2pt of specbar.east, anchor=west] {\speclevel{3}};
    \node[font=\tiny, text=black!50, below=2pt of specbar] {Specification Level};

    % ── SINGLE LANE (top, y=1.55) ──
    \def\singley{1.55}
    \def\firstcol{3.6}

    % Pipeline nodes
    \node[bluebox, minimum width=2.0cm]
          at (\firstcol, \singley) (single)
         {\textbf{Single Agent}\\[-2pt]{\tiny\ttfamily \_\_init\_\_ visible}};
    \node[graybox, right=0.7cm of single] (classS) {Complete\\[-2pt]Class};
    \node[graybox, right=0.7cm of classS] (testS) {Test Suite};
    \node[bluebox, right=0.7cm of testS] (passS) {Pass Rate};

    % Lane background
    \begin{scope}[on background layer]
      \fill[colSingle!5, lane]
        ([xshift=-0.25cm, yshift=0.5cm]single.north west)
        rectangle ([xshift=0.25cm, yshift=-0.5cm]passS.south east);
    \end{scope}

    % Lane label (inside background, left edge)
    \node[font=\scriptsize\bfseries, colSingle, anchor=east]
          at ([xshift=-0.35cm]single.west |- single) {Single};

    % Arrows — skeleton to single
    \draw[arr] ([yshift=0.3cm]skel.east) -- (single.west);
    \draw[arr] (single) -- (classS);
    \draw[arr] (classS) -- (testS);
    \draw[arr] (testS) -- (passS);

    % ── SPLIT LANE (bottom, y=-1.05) ──
    \def\splity{-1.05}

    % Agent boxes — centered vertically on splity
    \node[orangebox, minimum width=1.8cm]
          at (\firstcol, \splity+0.55) (agentA)
         {\textbf{Agent A}\\[-2pt]{\tiny list bias}};
    \node[orangebox, minimum width=1.8cm]
          at (\firstcol, \splity-0.55) (agentB)
         {\textbf{Agent B}\\[-2pt]{\tiny dict bias}};

    % Merge — centered between agents, at same x as classS
    \node[graybox] at (classS.center |- 0,\splity) (merge)
         {Na\"ive\\[-2pt]Merge};
    \node[graybox, right=0.7cm of merge] (classM) {Merged\\[-2pt]Class};
    \node[graybox, right=0.7cm of classM] (testM) {Test Suite};
    \node[orangebox, right=0.7cm of testM] (passM) {Pass Rate};

    % Lane background
    \begin{scope}[on background layer]
      \fill[colSplit!5, lane]
        ([xshift=-0.25cm, yshift=0.5cm]agentA.north west)
        rectangle ([xshift=0.25cm, yshift=-0.5cm]passM.south east);
    \end{scope}

    % Lane label
    \node[font=\scriptsize\bfseries, colSplit, anchor=east]
          at ([xshift=-0.35cm]agentA.west |- 0,\splity) {Split};

    % Arrows — skeleton to agents (fork)
    \draw[arr] ([yshift=-0.3cm]skel.east) -- ++(0.5cm,0)
              coordinate (splitpt) |- (agentA.west);
    \draw[arr] (splitpt) |- (agentB.west);

    % Arrows — agents to merge (simple |- routing)
    \draw[arr] (agentA.east) -- ++(0.25cm,0) |- (merge.north west);
    \draw[arr] (agentB.east) -- ++(0.25cm,0) |- (merge.south west);

    % Arrows — pipeline
    \draw[arr] (merge) -- (classM);
    \draw[arr] (classM) -- (testM);
    \draw[arr] (testM) -- (passM);

    % ── AST Detector (below split lane) ──
    \node[box, fill=colRecall!6, draw=colRecall!50, dashed,
          minimum width=1.6cm, font=\small,
          below=0.4cm of agentB] (detector) {AST Detector};
    % Agent A arrow: exit left side, route down outside Agent B, then into detector
    \draw[darr] ([yshift=-0.15cm]agentA.west) -- ++(-0.35cm, 0) |- (detector.west);
    % Agent B arrow: straight down
    \draw[darr] (agentB.south) -- (detector.north);
    \node[font=\tiny, text=colRecall!70, right=0.1cm of detector.east,
          anchor=west] {$\rightarrow$ conflicts};

  \end{tikzpicture}
  \caption{Experimental overview. Each task's class skeleton is processed at specification level $\ell\in\{$\speclevel{0}\textendash\speclevel{3}$\}$. The \emph{single} condition (top) gives one unbiased agent the full skeleton including \texttt{\_\_init\_\_}. The \emph{split} condition (bottom) hides \texttt{\_\_init\_\_} and assigns methods to two biased agents whose outputs are merged. An AST-based detector (dashed) checks for structural conflicts before integration.}
  \label{fig:overview}
\end{figure*}

\Cref{fig:overview} summarizes the full experimental pipeline, from class skeleton to agent generation, conflict analysis, and downstream evaluation.

\subsection{Tasks and Benchmark}

We use 51 tasks from ClassEval~\citep{du2024classeval}, each with $\geq$3 methods (from 88 eligible tasks in the full ClassEval set of 100). For each task, we generate four specification variants (\speclevel{0}--\speclevel{3}) using progressive information removal. The resulting benchmark, \textbf{AmbigClass}, contains 204 task-specification pairs. The experiment requires 612 LLM calls (3 per task-level: 1 single + 2 split agents) and 408 test suite executions.

\subsection{Agent Configuration}

Each configuration choice is motivated by the need to create a controlled setting in which specification-driven coordination failures arise reliably while remaining representative of real multi-agent workflows.

\begin{itemize}
  \item \textbf{Model}: Claude Sonnet 4.\footnote{API model key: \texttt{claude-sonnet-4-20250514}} We use the same model for all agents to ensure that performance differences reflect specification and coordination effects rather than capability differences between models.
  \item \textbf{Temperature}: 0.7 for biased agents, 0.0 for single agent. The elevated temperature for split agents increases output diversity, making it more likely that latent specification ambiguities surface as observable conflicts. The single agent uses greedy decoding to provide a stable, reproducible ceiling.
  \item \textbf{Biases}: Agent~A's system prompt includes ``always store data as LISTS''; Agent~B includes ``always store data as DICTS.'' This deliberate bias serves as a controlled proxy for the implicit stylistic divergences that emerge when different developers---or agents---make independent design choices without a shared data-structure contract. The list-vs-dictionary opposition was chosen because it produces structurally incompatible code that is easy to detect via AST analysis yet representative of real-world type conflicts.
  \item \textbf{Task split}: Methods are divided into two disjoint groups by alternating index. This simple, deterministic scheme guarantees that both agents contribute to every task and that inter-method dependencies are spread across agents rather than concentrated within one, maximizing the coordination surface.
\end{itemize}

\subsection{Evaluation}

Each condition's output is evaluated against the original ClassEval test suite. We report the proportion of tests passed (0--100\%). A task is considered ``successful'' if $\geq$80\% of tests pass.

\subsection{Recovery Experiment (RQ4)}

To isolate the contributions of specification quality and conflict information to integration recovery, we conduct a $2\times 2$ factorial experiment on 53 tasks. Two biased agents always generate code under \speclevel{3} specifications (bare signatures). Their outputs are then given to a \emph{merger agent} that attempts to combine them into a correct class. The factorial manipulates what the merger receives:

\begin{description}
  \item[Specification level] Either \speclevel{3} (the same minimal spec the agents used) or \speclevel{0} (the full specification with docstrings and data-structure references).
  \item[Conflict report] Either absent or present (the AST detector's report of type, state, and protocol conflicts between the two agents' outputs).
\end{description}

This yields four merge conditions---\emph{Blind} (\speclevel{3}, no report), \emph{Guided} (\speclevel{3}, with report), \emph{Spec-Only} (\speclevel{0}, no report), and \emph{Resolve} (\speclevel{0}, with report)---plus two baselines: \emph{Single} (one unbiased agent with \speclevel{0} and \texttt{\_\_init\_\_} visible) and \emph{Na\"ive} (direct concatenation without any merger). Total cost: \$3.54.

In total, the experimental design produces 204 task-specification pairs for the main experiment and 318 merger runs for the recovery experiment, enabling a systematic comparison across specification levels, agent conditions, and recovery strategies. We now turn to the results.

% ============================================================
\section{Results}
\label{sec:results}

We present the results organized by research question. First, we show that specification completeness is the dominant predictor of multi-agent integration success (RQ1). We then quantify the coordination penalty that persists even under full specifications (RQ2), demonstrate that AST-based conflict detection reliably signals specification inadequacy (RQ3), and finally evaluate which factors---specification quality, conflict information, or their combination---enable recovery from integration failures (RQ4).

\subsection{RQ1: Specification Completeness Predicts Integration Success}

RQ1 asks whether the amount of structural information in the specification predicts whether independently generated agent outputs can be successfully integrated. We compare Single and Split conditions across all four specification levels, measuring test pass rates and the number of AST-detected conflicts.

\begin{table}[h]
  \caption{Test pass rates by specification level and condition ($n=51$ tasks). Conflicts = mean AST-detected conflicts per task.}
  \label{tab:main-results}
  \centering
  \begin{tabular}{lcccc}
    \toprule
    Level & Single (\%) & Split (\%) & Gap (pp) & Conflicts (mean) \\
    \midrule
    \speclevel{0} & 88.6 & 58.2 & +30.4 & 1.0 \\
    \speclevel{1} & 77.8 & 43.0 & +34.8 & 1.4 \\
    \speclevel{2} & 66.0 & 36.5 & +29.5 & 1.5 \\
    \speclevel{3} & 55.8 & 24.6 & +31.3 & 1.6 \\
    \midrule
    Mean & 72.1 & 40.6 & +31.5 & 1.4 \\
    \bottomrule
  \end{tabular}
\end{table}

\begin{figure}[t]
  \centering
  \begin{tikzpicture}
    \begin{axis}[
      width=9cm, height=6.5cm,
      axis lines=left,
      xlabel={Specification Level},
      ylabel={Test Pass Rate (\%)},
      xtick={0,1,2,3},
      xticklabels={\speclevel{0},\speclevel{1},\speclevel{2},\speclevel{3}},
      ymin=15, ymax=100,
      ytick={20,30,40,50,60,70,80,90,100},
      ymajorgrids=true,
      grid style={dashed, black!20},
      legend style={at={(0.98,0.98)}, anchor=north east,
                    font=\small, draw=none, fill=white, fill opacity=0.8},
      every axis plot/.append style={line width=1.2pt, mark size=3pt},
      clip=false,
    ]
      % Fill-between for the gap
      \addplot[name path=single, colSingle, mark=*, mark options={solid}]
        coordinates {(0,88.6) (1,77.8) (2,66.0) (3,55.8)};
      \addplot[name path=split, colSplit, mark=square*, mark options={solid}]
        coordinates {(0,58.2) (1,43.0) (2,36.5) (3,24.6)};
      \addplot[colGap, fill opacity=0.15] fill between[of=single and split];

      % Value labels — shifted right at L0 to avoid y-axis overlap
      \node[above right, font=\scriptsize, colSingle, inner sep=1.5pt] at (axis cs:0,88.6) {88.6};
      \node[above, font=\scriptsize, colSingle] at (axis cs:1,77.8) {77.8};
      \node[above, font=\scriptsize, colSingle] at (axis cs:2,66.0) {66.0};
      \node[above, font=\scriptsize, colSingle] at (axis cs:3,55.8) {55.8};
      \node[below right, font=\scriptsize, colSplit!70!black, inner sep=1.5pt] at (axis cs:0,58.2) {58.2};
      \node[below, font=\scriptsize, colSplit!70!black] at (axis cs:1,43.0) {43.0};
      \node[below, font=\scriptsize, colSplit!70!black] at (axis cs:2,36.5) {36.5};
      \node[below, font=\scriptsize, colSplit!70!black] at (axis cs:3,24.6) {24.6};

      % Gap annotation
      \draw[{Stealth[length=4pt]}-{Stealth[length=4pt]}, colGap, line width=0.6pt]
        (axis cs:1.15,43.0) -- (axis cs:1.15,77.8)
        node[midway, right, font=\scriptsize\bfseries, colGap] {25--39pp};

      % Dashed vertical line: data-structure refs removed
      \draw[dashed, black!40, line width=0.6pt]
        (axis cs:1.5,15) -- (axis cs:1.5,98);
      \node[font=\tiny, text=black!50, rotate=90, anchor=south]
        at (axis cs:1.5,97) {Data-structure refs removed};

      \legend{Single, Split}
    \end{axis}
  \end{tikzpicture}
  \caption{Test pass rates degrade monotonically as specification detail is removed. The shaded band highlights the persistent coordination gap (25--39pp across models) between single-agent and split-agent conditions. The dashed line marks the transition where explicit data-structure references are stripped (\speclevel{1}$\to$\speclevel{2}).}
  \label{fig:degradation}
\end{figure}
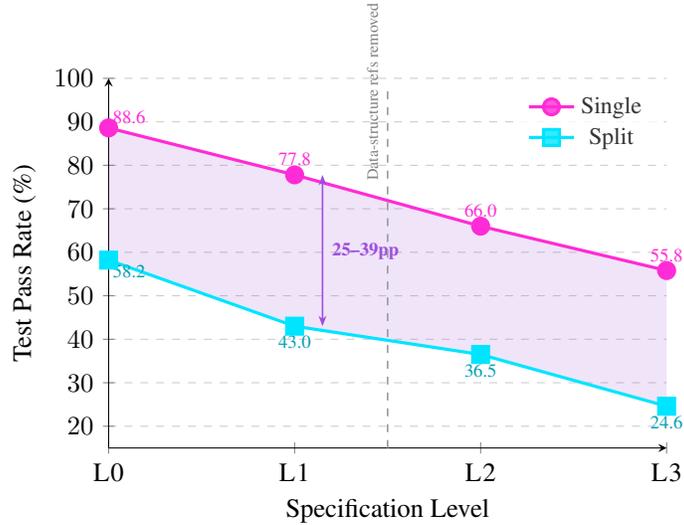

Both conditions degrade monotonically as specification detail is removed (\Cref{tab:main-results,fig:degradation}), confirming that specification completeness is a strong predictor of integration success. At \speclevel{0}, where docstrings explicitly name data structures and include usage examples, the single agent achieves 88.6\% and the split condition reaches 58.2\%. By \speclevel{3}, where only bare method signatures remain, these figures drop to 55.8\% and 24.6\%, respectively. The overall magnitude of degradation is similar for both conditions ($\sim$33pp), yet the split condition starts from a substantially lower baseline because coordination errors compound information loss: even at \speclevel{0}, agents that cannot see the shared \texttt{\_\_init\_\_} body must independently infer compatible data structures from the specification alone. The mean number of AST-detected conflicts also rises with specification degradation (1.0 at \speclevel{0} to 1.6 at \speclevel{3}), providing an independent structural signal that corroborates the test-based results.

\subsection{RQ2: The Coordination Tax}

RQ2 investigates whether the performance gap between single-agent and multi-agent conditions can be closed by improving specification quality, or whether multi-agent coordination imposes an irreducible cost. If specifications were the only bottleneck, we would expect the gap to vanish at \speclevel{0}.

The gap between Single and Split is significant at every level, ranging from 29.5 to 34.8pp for Sonnet and 25.5 to 38.8pp for Haiku (\Cref{tab:replication}). This ``coordination tax'' is \emph{not} eliminated by better specifications---even at \speclevel{0}, where docstrings explicitly name data structures, the two-agent condition loses 25--30 percentage points.

Wilcoxon signed-rank tests confirm the gap is significant at every level (all $p < 0.001$; Cohen's $d = 0.71$--$1.08$). A Friedman test finds no significant variation in gap magnitude across levels ($\chi^2 = 6.93$, $p = 0.074$), indicating that the coordination tax neither increases nor decreases with specification degradation. This finding has a direct practical implication: even when teams invest in comprehensive specifications, multi-agent architectures that divide code generation across agents will still incur a measurable performance penalty relative to a single agent that sees the full context. The penalty stems not from ambiguous specifications but from the inherent difficulty of producing structurally compatible code without shared state.

\subsection{RQ3: Conflict Detection as a Specification Signal}

RQ3 examines whether AST-based conflict detection can serve as an automated diagnostic for specification inadequacy. If structural conflicts between agent outputs correlate with specification degradation and predict integration failure, the detector would provide an actionable signal---without requiring any additional LLM calls---that current specifications are insufficient for reliable multi-agent coordination.

Conflicts increase monotonically with specification degradation (mean 1.0 to 1.6 per task; see \Cref{fig:conflicts} for the full taxonomy). Type conflicts dominate at every level (60.6\% of all 284 detections), growing fastest as data-structure references are removed.

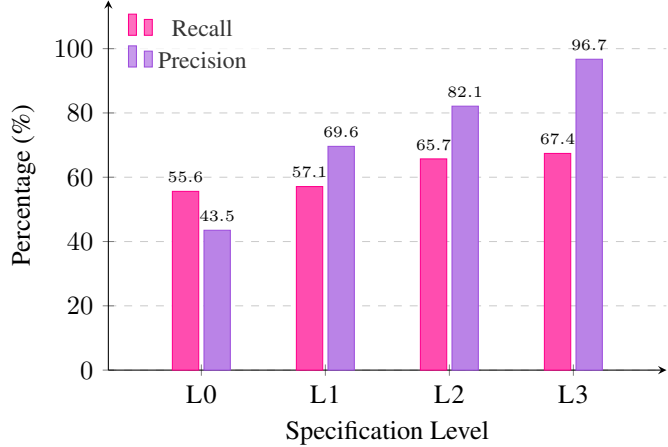
\begin{figure}[t]
  \centering
  \begin{tikzpicture}
    \begin{axis}[
      width=9cm, height=6.5cm,
      ybar,
      bar width=0.35cm,
      axis lines=left,
      xlabel={Specification Level},
      ylabel={Percentage (\%)},
      xtick={0,1,2,3},
      xticklabels={\speclevel{0},\speclevel{1},\speclevel{2},\speclevel{3}},
      ymin=0, ymax=115,
      ytick={0,20,40,60,80,100},
      ymajorgrids=true,
      grid style={dashed, black!20},
      legend style={at={(0.02,0.98)}, anchor=north west,
                    font=\small, draw=none, fill=white, fill opacity=0.8},
      nodes near coords,
      every node near coord/.append style={font=\tiny, anchor=south},
      enlarge x limits=0.25,
    ]
      \addplot[fill=colRecall!70, draw=colRecall]
        coordinates {(0,55.6) (1,57.1) (2,65.7) (3,67.4)};
      \addplot[fill=colPrecision!70, draw=colPrecision]
        coordinates {(0,43.5) (1,69.6) (2,82.1) (3,96.7)};
      \legend{Recall, Precision}
    \end{axis}
  \end{tikzpicture}
  \caption{AST conflict detector recall and precision by specification level. Precision improves dramatically as specifications degrade, reaching 96.7\% at \speclevel{3}---the detector is most reliable precisely where it is most needed.}
  \label{fig:detector}
\end{figure}

The detector operates at zero inference cost: it analyzes only the generated ASTs and requires no additional LLM calls. \Cref{tab:detector,fig:detector} show its performance. Overall recall is 62.9\% and precision is 75.0\%, but these averages hide the main pattern. Precision rises from 43.5\% at \speclevel{0} to \textbf{96.7\%} at \speclevel{3}, where weak specifications produce structural conflicts rather than subtler semantic incompatibilities.

\begin{table}[h]
  \caption{AST conflict detector performance by specification level. A task is ``failed'' if Split pass rate $<$50\%; ``detected'' if $\geq$1 conflict found.}
  \label{tab:detector}
  \centering
  \begin{tabular}{lcccccc}
    \toprule
    Level & $n$ & TP & FN & FP & Recall & Precision \\
    \midrule
    \speclevel{0} & 51 & 10 & 8 & 13 & 55.6\% & 43.5\% \\
    \speclevel{1} & 51 & 16 & 12 & 7 & 57.1\% & 69.6\% \\
    \speclevel{2} & 51 & 23 & 12 & 5 & 65.7\% & 82.1\% \\
    \speclevel{3} & 51 & 29 & 14 & 1 & 67.4\% & \textbf{96.7\%} \\
    \midrule
    All   & 204 & 78 & 46 & 26 & 62.9\% & 75.0\% \\
    \bottomrule
  \end{tabular}
\end{table}

A breakdown by conflict type (\Cref{fig:conflicts} in \Cref{app:taxonomy}) confirms that type conflicts dominate at all levels, growing from 26 at \speclevel{0} to 53 at \speclevel{3}; state conflicts are second, while protocol and return conflicts remain rare.

At \speclevel{0}, false positives arise because structural differences do not always cause test failures---agents may use different internal representations that happen to satisfy the same interface contract. False negatives reflect the complementary limitation: semantically incorrect but structurally compatible code that AST analysis cannot catch, such as two agents that both use dictionaries but disagree on key names.

Together, these results establish the AST detector as a useful---though imperfect---specification adequacy signal. Its high precision at weak specification levels means that when it flags a conflict, the specification almost certainly lacks the structural detail needed for reliable coordination. The detector can \emph{diagnose} integration failures---but can conflict information help a merger agent \emph{fix} them?

\subsection{RQ4: Recovery Requires Specification, Not Conflict Reports}

RQ4 asks what enables a merger agent to recover from integration failures produced by biased split agents: richer specifications, conflict reports, or their combination. This question has direct practical significance---it determines whether teams should invest in building better conflict-detection infrastructure or in improving the specifications that agents receive. The $2\times 2$ factorial design (\Cref{sec:design}) independently varies specification level (\speclevel{3} vs.\ \speclevel{0}) and conflict report availability (absent vs.\ present), yielding four merge conditions tested on 53 tasks.

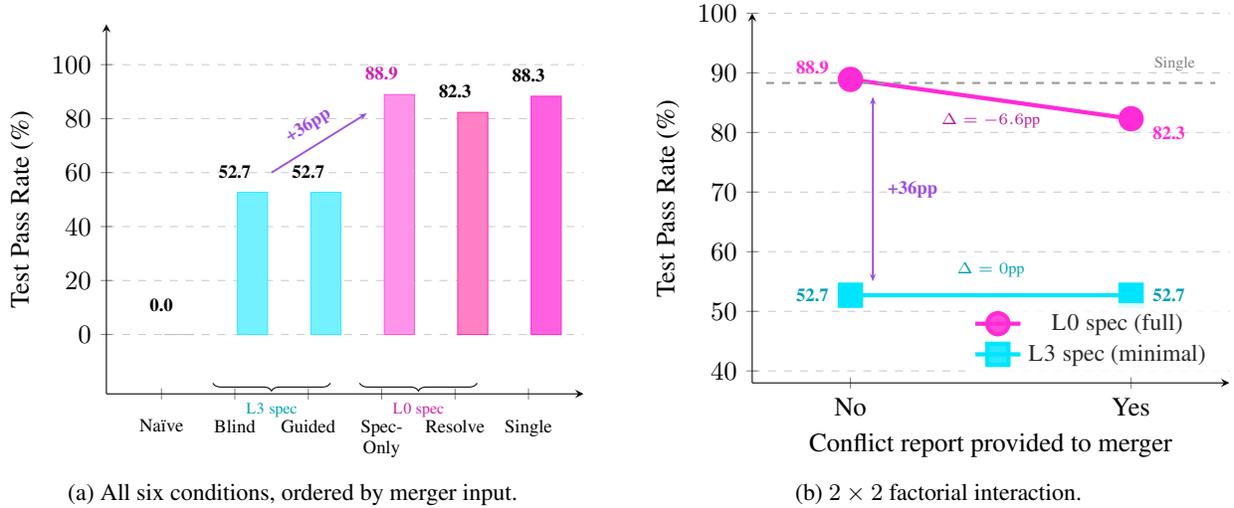
\begin{figure*}[t]
  \centering
  \begin{subfigure}[b]{0.48\textwidth}
    \centering
    \begin{tikzpicture}
      \begin{axis}[
        ybar, bar width=0.40cm,
        width=\textwidth, height=6.5cm,
        axis lines=left,
        ylabel={Test Pass Rate (\%)},
        ymin=-22, ymax=115,
        ytick={0,20,40,60,80,100},
        ymajorgrids=true,
        grid style={dashed, black!20},
        xtick={0,1,2,3,4,5},
        xticklabels={Na\"ive, Blind, Guided, {Spec-\\Only}, Resolve, Single},
        xticklabel style={font=\scriptsize, align=center, anchor=north},
        x tick label style={yshift=-4pt},
        enlarge x limits=0.15,
        clip=false,
      ]
        % Bars with per-condition colors
        \addplot[fill=colDeepBlack!8, draw=colDeepBlack!25, forget plot] coordinates {(0, 0)};
        \addplot[fill=colSplit!55, draw=colSplit!80, forget plot] coordinates {(1, 52.7)};
        \addplot[fill=colSplit!55, draw=colSplit!80, forget plot] coordinates {(2, 52.7)};
        \addplot[fill=colSingle!50, draw=colSingle!80, forget plot] coordinates {(3, 88.9)};
        \addplot[fill=colRecall!50, draw=colRecall!80, forget plot] coordinates {(4, 82.3)};
        \addplot[fill=colSingle!75, draw=colSingle, forget plot] coordinates {(5, 88.3)};
        % Value labels
        \node[above=2pt, font=\scriptsize\bfseries] at (axis cs:0,3) {0.0};
        \node[above=2pt, font=\scriptsize\bfseries] at (axis cs:1,52.7) {52.7};
        \node[above=2pt, font=\scriptsize\bfseries] at (axis cs:2,52.7) {52.7};
        \node[above=2pt, font=\scriptsize\bfseries, colSingle!80!black] at (axis cs:3,88.9) {88.9};
        \node[above=2pt, font=\scriptsize\bfseries] at (axis cs:4,82.3) {82.3};
        \node[above=2pt, font=\scriptsize\bfseries] at (axis cs:5,88.3) {88.3};
        % Grouping braces (below tick labels)
        \draw[decorate, decoration={brace, amplitude=3pt, mirror}]
          (axis cs:0.7,-18) -- (axis cs:2.3,-18)
          node[midway, below=4pt, font=\tiny, colSplit!70!black] {\speclevel{3} spec};
        \draw[decorate, decoration={brace, amplitude=3pt, mirror}]
          (axis cs:2.7,-18) -- (axis cs:4.3,-18)
          node[midway, below=4pt, font=\tiny, colSingle!80!black] {\speclevel{0} spec};
        % Spec effect arrow
        \draw[-{Stealth[length=3pt]}, colGap, line width=0.7pt]
          (axis cs:1.5, 60) -- (axis cs:2.8, 82)
          node[midway, above=1pt, font=\scriptsize\bfseries, colGap, sloped] {+36pp};
      \end{axis}
    \end{tikzpicture}
    \caption{All six conditions, ordered by merger input.}
    \label{fig:recovery-bars}
  \end{subfigure}
  \hfill
  \begin{subfigure}[b]{0.48\textwidth}
    \centering
    \begin{tikzpicture}
      \begin{axis}[
        width=\textwidth, height=6.5cm,
        axis lines=left,
        xlabel={Conflict report provided to merger},
        ylabel={Test Pass Rate (\%)},
        xtick={0,1},
        xticklabels={No, Yes},
        ymin=38, ymax=100,
        ytick={40,50,60,70,80,90,100},
        ymajorgrids=true,
        grid style={dashed, black!20},
        legend style={at={(0.98,0.02)}, anchor=south east,
                      font=\small, draw=none, fill=white, fill opacity=0.85},
        every axis plot/.append style={line width=1.6pt, mark size=4pt},
        clip=false,
        xmin=-0.35, xmax=1.35,
      ]
        % L0 spec line (plum)
        \addplot[colSingle, mark=*, mark options={solid, fill=colSingle}]
          coordinates {(0,88.9) (1,82.3)};
        % L3 spec line (pink)
        \addplot[colSplit, mark=square*, mark options={solid, fill=colSplit}]
          coordinates {(0,52.7) (1,52.7)};
        % Single baseline (dashed)
        \draw[dashed, black!40, line width=0.8pt]
          (axis cs:-0.3,88.3) -- (axis cs:1.3,88.3);
        \node[font=\tiny, black!50, anchor=south west] at (axis cs:1.05,88.8) {Single};
        % Value labels
        \node[left=7pt, font=\scriptsize\bfseries, colSingle, inner sep=1pt]
          at (axis cs:0,91) {88.9};
        \node[right=7pt, font=\scriptsize\bfseries, colSingle, inner sep=1pt]
          at (axis cs:1,80) {82.3};
        \node[left=7pt, font=\scriptsize\bfseries, colSplit!70!black, inner sep=1pt]
          at (axis cs:0,52.7) {52.7};
        \node[right=7pt, font=\scriptsize\bfseries, colSplit!70!black, inner sep=1pt]
          at (axis cs:1,52.7) {52.7};
        % Spec effect arrow (left side)
        \draw[{Stealth[length=3pt]}-{Stealth[length=3pt]}, colGap, line width=0.7pt]
          (axis cs:0.08,55) -- (axis cs:0.08,86)
          node[midway, right=2pt, font=\scriptsize\bfseries, colGap] {+36pp};
        % Delta annotations
        \node[font=\tiny, colSplit!70!black] at (axis cs:0.5,57) {$\Delta = 0$pp};
        \node[font=\tiny, colSingle!70!black]
          at (axis cs:0.5,82) {$\Delta = -6.6$pp};
        \legend{\speclevel{0} spec (full), \speclevel{3} spec (minimal)}
      \end{axis}
    \end{tikzpicture}
    \caption{$2\times 2$ factorial interaction.}
    \label{fig:factorial}
  \end{subfigure}
  \caption{Recovery experiment results ($n=53$ tasks). (a)~Six conditions showing Spec-Only (88.9\%) matches Single (88.3\%), while Blind and Guided are identical (52.7\%)---conflict information has no effect without the full specification. (b)~Interaction plot: the \speclevel{0} specification accounts for +36pp recovery; conflict reports contribute $\Delta = 0$pp at \speclevel{3} and $-6.6$pp at \speclevel{0}. Dashed line: single-agent baseline.}
  \label{fig:recovery}
\end{figure*}

\Cref{tab:recovery,fig:recovery,fig:recovery-bars,fig:factorial} make the result unambiguous. The specification effect dominates (+36.2pp). Conflict information contributes nothing at \speclevel{3} ($\Delta = 0$pp) and slightly hurts at \speclevel{0} ($-6.6$pp).

\begin{table}[h]
  \caption{Recovery experiment: $2\times 2$ factorial ($n=53$ tasks). 120 AST-detected conflicts (79 type, 37 state, 3 protocol, 1 return).}
  \label{tab:recovery}
  \centering
  \begin{tabular}{llcc}
    \toprule
    Condition & Spec & Conflicts & Pass rate (\%) \\
    \midrule
    Single (ceiling) & \speclevel{0} & --- & 88.3 \\
    Na\"ive (floor) & --- & --- & 0.0 \\
    \midrule
    Blind            & \speclevel{3} & No  & 52.7 \\
    Guided           & \speclevel{3} & Yes & 52.7 \\
    Spec-Only        & \speclevel{0} & No  & 88.9 \\
    Resolve          & \speclevel{0} & Yes & 82.3 \\
    \midrule
    \multicolumn{4}{l}{\footnotesize Spec effect: $+36.2$pp \quad Conflict effect: $+0.0$pp / $-6.6$pp \quad Interaction: $-6.6$pp} \\
    \bottomrule
  \end{tabular}
\end{table}

The critical finding is the Spec-Only condition. At 88.9\%, it matches---and even marginally exceeds---the single-agent ceiling (88.3\%), showing that the full specification alone is sufficient to recover from integration failure. This means that a merger agent equipped with a complete specification can reconstruct a correct implementation from two structurally incompatible agent outputs, effectively undoing the coordination damage caused by partial knowledge.

Equally striking is the comparison between Blind and Guided (both 52.7\%): adding conflict reports to a minimal specification produces zero improvement. Without the structural context that the full specification provides, the merger agent cannot act on the detector's diagnoses---knowing \emph{that} two agents disagree on a data structure is useless without knowing \emph{what} the correct structure should be.

The negative interaction ($-6.6$pp in the Resolve condition) suggests that conflict reports introduce noise when the specification already resolves the structural mismatch. One plausible mechanism is that the detailed conflict report biases the merger toward local, symptom-level fixes rather than the holistic redesign that the full specification enables, paralleling findings where redundant information degrades human decision-making~\citep{engler2001bugs}.

% ============================================================
\section{Discussion}
\label{sec:discussion}

The results in \Cref{sec:results} establish three core findings: specification completeness predicts integration success, a persistent coordination tax survives even under full specifications, and recovery from integration failure is driven almost entirely by specification quality rather than conflict diagnostics. In this section, we interpret these findings through the lens of classical software engineering principles, examine the sources of task-level variability, disentangle coordination cost from information asymmetry through an additional factorial experiment, and address the limitations of our experimental design. Together, these analyses clarify both the practical implications of the specification gap and the boundaries of our claims.

\subsection{Specifications as Coordination and Recovery}

Our results lend empirical support to an intuition from classical software engineering: specifications are not merely documentation; they also function as \emph{coordination mechanisms}. When two code agents share a specification that says ``add to the \texttt{job\_listings} list,'' both produce list-based implementations regardless of their individual biases. When the specification says only ``Publish positions,'' each agent falls back to its prior and integration fails. In this sense, our findings align with Meyer's Design by Contract~\citep{meyer1992dbc}: for code agents operating under partial knowledge, ``precise enough'' appears to require data-structure information, not only behavioral descriptions.

The factorial recovery experiment (\Cref{tab:recovery}) qualifies this interpretation. In our setting, the AST conflict detector is valuable as a \emph{diagnostic} tool---it helps explain \emph{why} integration fails---but it does not improve recovery once a merger agent is involved. By contrast, the full specification is sufficient to restore performance to the single-agent ceiling. One plausible interpretation is that the specification already resolves the structural ambiguities that the conflict reports expose, leaving little additional value for the reports themselves.

The practical implication is correspondingly cautious: when code agents fail to integrate under conditions similar to ours---shared-class generation with opposing structural biases---improving the specification is likely to be more effective than adding conflict-detection infrastructure alone.

A natural concern is whether the prompt-injected biases in our design overstate the coordination problem. We argue that the injection models a realistic operational risk: in practice, agents are configured with different system prompts, reuse prompts inherited from prior projects, or are trained on codebases with divergent stylistic norms. The list-vs-dictionary opposition was chosen because it produces structurally detectable conflicts, but the underlying mechanism---independent agents defaulting to incompatible priors under ambiguous specifications---does not require this level of explicitness to operate. The init-visibility experiment (\Cref{sec:init-visibility}) supports this interpretation: even when both agents share the identical constructor body and are instructed to preserve it, the coordination tax persists at +15.7pp, indicating that the gap is not an artifact of the bias injection but a property of independent generation itself.

\subsection{The Coordination Tax and Task-Level Variability}

Even with full specifications (\speclevel{0}), multi-agent integration loses 25--30pp relative to a single agent. This gap persists at every specification level (25--39pp across two models), suggesting a coordination cost that specification quality alone does not eliminate in our setting. Part of the gap does stem from a design asymmetry: single agents see the \texttt{\_\_init\_\_} body, while split agents must infer data structures. The init-visibility experiment (\Cref{sec:init-visibility}) separates these effects and indicates that coordination is the \emph{larger} component (+16--20pp), while information asymmetry is secondary (+11--15pp); the two are approximately additive.

The aggregate gap conceals substantial task-level variation (std 27--43pp). Method count does not predict the coordination gap ($r \approx 0$, $p > 0.5$ for both models). Instead, task difficulty depends on whether agents' biases happen to be compatible. Cross-model category agreement is 72\%\ (36/50 common tasks), with ``easy'' tasks sharing a common trait: simple, unambiguous state where list-vs-dictionary bias is irrelevant.

In a minority of cases, split agents actually outperform the single agent (13.7\% of tasks at \speclevel{0}, 5.9\% at \speclevel{3}). This occurs when the biased agents' forced data structure accidentally matches the expected interface better than the single agent's unconstrained choice.

\subsection{Disentangling Coordination from Information Asymmetry}
\label{sec:init-visibility}

The preceding gap conflates two factors: \emph{coordination cost} (two biased agents must agree on data structures) and \emph{information asymmetry} (split agents do not see the \texttt{\_\_init\_\_} body that single agents receive). To separate these, we run a $2 \times 2$ factorial experiment ($n=50$ tasks) crossing agent mode (single vs.\ split) with \texttt{\_\_init\_\_} visibility (visible vs.\ hidden), creating four conditions (see \Cref{app:init-visibility} for full prompts). \Cref{tab:init-visibility} reports the results.

\begin{table}[h]
  \caption{Init visibility experiment: $2\times2$ factorial ($n=50$ tasks). Top: mean pass rates (\%) across all spec levels. Bottom: per-level detail.}
  \label{tab:init-visibility}
  \centering
  \begin{tabular}{lccc}
    \toprule
    & Init Visible (\%) & Init Hidden (\%) & Info Effect (pp) \\
    \midrule
    Single       & 72.3 & 61.1 & +11.2 \\
    Split        & 56.6 & 41.2 & +15.3 \\
    Coord Effect (pp) & +15.7 & +19.8 & Interaction: $-4.1$ \\
    \bottomrule
  \end{tabular}

  \vspace{6pt}
  \centering\small
  \begin{tabular}{lcccccccc}
    \toprule
    & \multicolumn{2}{c}{Single (\%)} & \multicolumn{2}{c}{Split (\%)} & \multicolumn{2}{c}{Info Effect} & \multicolumn{2}{c}{Coord Effect} \\
    \cmidrule(lr){2-3} \cmidrule(lr){4-5} \cmidrule(lr){6-7} \cmidrule(lr){8-9}
    Level & Vis & Hid & Vis & Hid & Single & Split & Vis & Hid \\
    \midrule
    \speclevel{0} & 88.4 & 81.1 & 70.8 & 59.3 & +7.3 & +11.5 & +17.6 & +21.8 \\
    \speclevel{1} & 77.4 & 63.5 & 58.5 & 43.5 & +13.9 & +15.0 & +18.9 & +20.0 \\
    \speclevel{2} & 66.9 & 54.8 & 53.4 & 37.2 & +12.1 & +16.3 & +13.5 & +17.6 \\
    \speclevel{3} & 56.5 & 44.8 & 43.5 & 24.9 & +11.7 & +18.5 & +13.0 & +19.9 \\
    \bottomrule
  \end{tabular}
\end{table}

\begin{figure}[t]
  \centering
  \begin{tikzpicture}
    \begin{axis}[
      width=9cm, height=6cm,
      ybar,
      bar width=0.35cm,
      axis lines=left,
      xlabel={Specification Level},
      ylabel={Effect Size (pp)},
      xtick={0,1,2,3},
      xticklabels={\speclevel{0},\speclevel{1},\speclevel{2},\speclevel{3}},
      ymin=0, ymax=28,
      ytick={0,5,10,15,20,25},
      ymajorgrids=true,
      grid style={dashed, black!20},
      legend style={at={(0.98,0.98)}, anchor=north east,
                    font=\small, draw=none, fill=white, fill opacity=0.85},
      nodes near coords,
      every node near coord/.append style={font=\tiny, anchor=south},
      enlarge x limits=0.25,
    ]
      % Coordination effect (mean of Vis and Hid coord columns)
      \addplot[fill=colSingle!60, draw=colSingle!80]
        coordinates {(0,19.7) (1,19.5) (2,15.6) (3,16.5)};
      % Information effect (mean of Single and Split info columns)
      \addplot[fill=colSplit!60, draw=colSplit!80]
        coordinates {(0,9.4) (1,14.5) (2,14.2) (3,15.1)};
      \legend{Coordination cost, Information asymmetry}
    \end{axis}
  \end{tikzpicture}
  \caption{Gap decomposition by specification level. The coordination effect (single vs.\ split, averaging over visibility) consistently exceeds the information effect (visible vs.\ hidden, averaging over agent mode) at every level. At \speclevel{0}, coordination accounts for twice the gap; by \speclevel{3}, the two components converge.}
  \label{fig:gap-decomposition}
\end{figure}
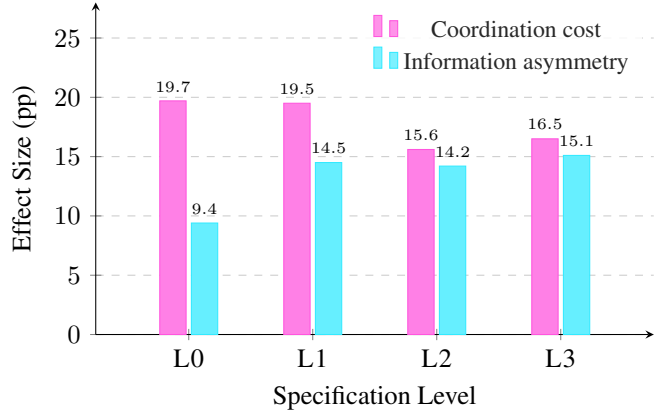

\textbf{Coordination is the larger factor.} Even when both agents see the identical \texttt{\_\_init\_\_} body and are told to preserve it, the list/dict bias still causes a +15.7pp penalty. This suggests that the specification gap is not merely an information confound. Across all settings, the coordination effect (+15.7 to +19.8pp) exceeds the information effect (+11.2 to +15.3pp).

\textbf{Information asymmetry is real but secondary.} Hiding \texttt{\_\_init\_\_} costs the single agent 11.2pp on average; the effect is larger for split agents (+15.3pp), where information loss compounds with coordination difficulty.

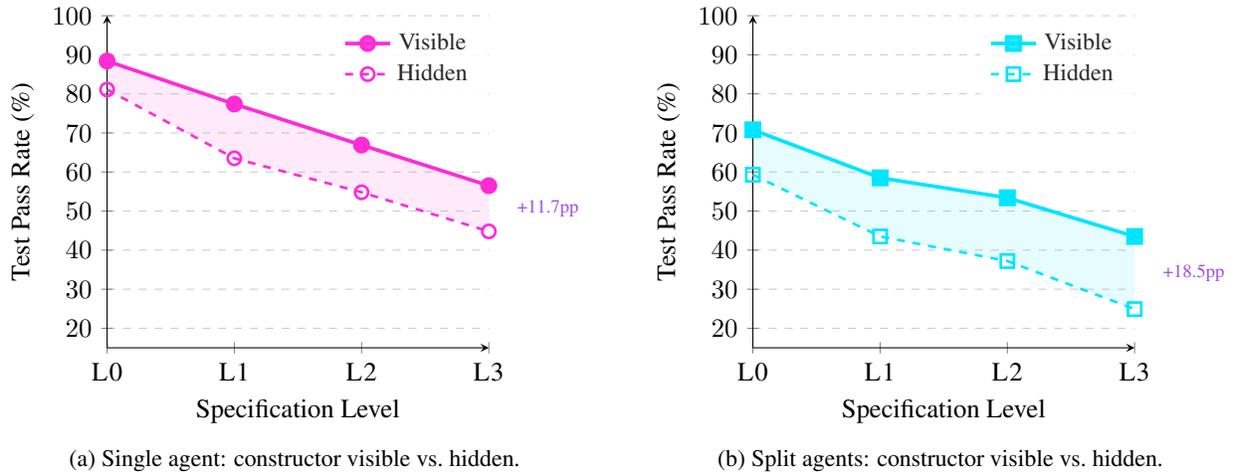
\begin{figure*}[t]
  \centering
  \begin{subfigure}[b]{0.48\textwidth}
    \centering
    \begin{tikzpicture}
      \begin{axis}[
        width=0.84\linewidth, height=6cm,
        axis lines=left,
        xlabel={Specification Level},
        ylabel={Test Pass Rate (\%)},
        xtick={0,1,2,3},
        xticklabels={\speclevel{0},\speclevel{1},\speclevel{2},\speclevel{3}},
        ymin=15, ymax=100,
        ytick={20,30,40,50,60,70,80,90,100},
        ymajorgrids=true,
        grid style={dashed, black!20},
        legend style={at={(0.98,0.98)}, anchor=north east,
                      font=\small, draw=none, fill=white, fill opacity=0.85},
        every axis plot/.append style={mark size=2.5pt},
        clip=false,
      ]
        \addplot[name path=svis, colSingle, line width=1.4pt, mark=*, mark options={solid, fill=colSingle}]
          coordinates {(0,88.4) (1,77.4) (2,66.9) (3,56.5)};
        \addplot[name path=shid, colSingle, line width=1.0pt, dashed, mark=o, mark options={solid, colSingle}]
          coordinates {(0,81.1) (1,63.5) (2,54.8) (3,44.8)};
        \addplot[colSingle, fill opacity=0.10] fill between[of=svis and shid];
        % Gap annotation
        \node[font=\scriptsize, colGap, anchor=west] at (axis cs:3.15,50.7) {+11.7pp};
        \legend{Visible, Hidden}
      \end{axis}
    \end{tikzpicture}
    \caption{Single agent: constructor visible vs.\ hidden.}
    \label{fig:init-vis-single}
  \end{subfigure}
  \hfill
  \begin{subfigure}[b]{0.48\textwidth}
    \centering
    \begin{tikzpicture}
      \begin{axis}[
        width=0.84\linewidth, height=6cm,
        axis lines=left,
        xlabel={Specification Level},
        ylabel={Test Pass Rate (\%)},
        xtick={0,1,2,3},
        xticklabels={\speclevel{0},\speclevel{1},\speclevel{2},\speclevel{3}},
        ymin=15, ymax=100,
        ytick={20,30,40,50,60,70,80,90,100},
        ymajorgrids=true,
        grid style={dashed, black!20},
        legend style={at={(0.98,0.98)}, anchor=north east,
                      font=\small, draw=none, fill=white, fill opacity=0.85},
        every axis plot/.append style={mark size=2.5pt},
        clip=false,
      ]
        \addplot[name path=spvis, colSplit, line width=1.4pt, mark=square*, mark options={solid, fill=colSplit}]
          coordinates {(0,70.8) (1,58.5) (2,53.4) (3,43.5)};
        \addplot[name path=sphid, colSplit, line width=1.0pt, dashed, mark=square, mark options={solid, colSplit}]
          coordinates {(0,59.3) (1,43.5) (2,37.2) (3,24.9)};
        \addplot[colSplit, fill opacity=0.10] fill between[of=spvis and sphid];
        % Gap annotation
        \node[font=\scriptsize, colGap, anchor=west] at (axis cs:3.15,34.2) {+18.5pp};
        \legend{Visible, Hidden}
      \end{axis}
    \end{tikzpicture}
    \caption{Split agents: constructor visible vs.\ hidden.}
    \label{fig:init-vis-split}
  \end{subfigure}
  \caption{Init visibility experiment. Solid lines: \texttt{\_\_init\_\_} visible; dashed lines: hidden. Shaded area: information effect (gap due to \texttt{\_\_init\_\_} visibility). The information gap widens as specifications degrade, especially for split agents (right), where it grows from +11.5pp at \speclevel{0} to +18.5pp at \speclevel{3}. The vertical distance between the two panels' solid lines represents the pure coordination effect.}
  \label{fig:init-visibility}
\end{figure*}

\textbf{The effects are approximately additive} (interaction $= -4.1$pp), meaning the gap decomposes reasonably cleanly into two mechanisms (\Cref{fig:gap-decomposition}). \Cref{fig:init-visibility} shows the per-level degradation curves, while \Cref{fig:init-vis-single,fig:init-vis-split} separate the same pattern for single-agent and split-agent settings. Notably, AST conflicts drop from 1.3 per task when \texttt{\_\_init\_\_} is hidden to 0.6 when it is visible, and they remain constant at 0.6 across all spec levels in the visible condition. This pattern suggests that the constructor body helps anchor agents to a shared data structure.

\subsection{Cross-Model Replication}

\begin{figure}[t]
  \centering
  \begin{tikzpicture}
    \begin{axis}[
      width=9cm, height=6.5cm,
      axis lines=left,
      xlabel={Specification Level},
      ylabel={Test Pass Rate (\%)},
      xtick={0,1,2,3},
      xticklabels={\speclevel{0},\speclevel{1},\speclevel{2},\speclevel{3}},
      ymin=15, ymax=100,
      ytick={20,30,40,50,60,70,80,90,100},
      ymajorgrids=true,
      grid style={dashed, black!20},
      legend style={at={(0.98,0.98)}, anchor=north east,
                    font=\small, draw=none, fill=white, fill opacity=0.85,
                    legend columns=1},
      every axis plot/.append style={mark size=2.5pt},
      clip=false,
    ]
      % Sonnet — solid, thicker
      \addplot[colSingle, line width=1.4pt, mark=*, mark options={solid, fill=colSingle}]
        coordinates {(0,88.6) (1,77.8) (2,66.0) (3,55.8)};
      \addplot[colSplit, line width=1.4pt, mark=square*, mark options={solid, fill=colSplit}]
        coordinates {(0,58.2) (1,43.0) (2,36.5) (3,24.6)};

      % Haiku — dashed, thinner, error bars from 3 runs
      \addplot[colSingle, line width=1.0pt, dashed, mark=o, mark options={solid, colSingle},
               error bars/.cd, y dir=both, y explicit, error bar style={line width=0.4pt, colSingle!60}]
        coordinates {(0,86.6) +- (0,0.1) (1,75.5) +- (0,0.3) (2,66.8) +- (0,0.6) (3,55.2) +- (0,0.7)};
      \addplot[colSplit, line width=1.0pt, dashed, mark=square, mark options={solid, colSplit},
               error bars/.cd, y dir=both, y explicit, error bar style={line width=0.4pt, colSplit!60}]
        coordinates {(0,61.1) +- (0,1.9) (1,36.7) +- (0,0.7) (2,31.8) +- (0,0.3) (3,25.4) +- (0,0.8)};

      \legend{Sonnet Single, Sonnet Split, Haiku Single, Haiku Split}
    \end{axis}
  \end{tikzpicture}
  \caption{Cross-model replication. Solid lines: Sonnet (1~run); dashed lines: Haiku (3-run mean, error bars show $\pm$1~std). Single-agent curves (pink) nearly overlap. Split-agent curves (cyan) follow the same degradation pattern. Error bars are barely visible, confirming high run-to-run stability.}
  \label{fig:replication}
\end{figure}
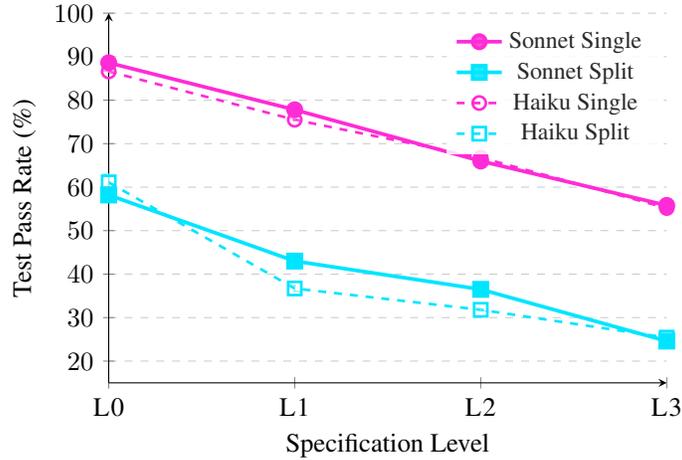

To test whether the specification gap is model-specific, we replicate the full experiment with Claude Haiku 4.5\footnote{API model key: \texttt{claude-haiku-4-5-20251001}}, a smaller and less capable model in the same family ($n=50$ tasks). This replication is intended as a robustness check on the phenomenon, not as a leaderboard comparison between Claude variants. To assess run-to-run stability, we execute the Haiku experiment three times (total cost: \$4.38 vs.\ \$4.83 for Sonnet). \Cref{tab:replication} reports Haiku values as mean $\pm$ standard deviation across the three runs, and \Cref{fig:replication} overlays the degradation curves.

\begin{table}[h]
  \caption{Cross-model replication. Sonnet ($n=51$, 1~run) vs.\ Haiku ($n=50$, 3~runs, mean $\pm$ std).}
  \label{tab:replication}
  \centering
  \begin{tabular}{lcccccc}
    \toprule
    & \multicolumn{2}{c}{Single (\%)} & \multicolumn{2}{c}{Split (\%)} & \multicolumn{2}{c}{Gap (pp)} \\
    \cmidrule(lr){2-3} \cmidrule(lr){4-5} \cmidrule(lr){6-7}
    Level & Sonnet & Haiku & Sonnet & Haiku & Sonnet & Haiku \\
    \midrule
    \speclevel{0} & 88.6 & 86.6{\tiny$\pm$0.1} & 58.2 & 61.1{\tiny$\pm$1.9} & +30.4 & +25.5{\tiny$\pm$1.8} \\
    \speclevel{1} & 77.8 & 75.5{\tiny$\pm$0.3} & 43.0 & 36.7{\tiny$\pm$0.7} & +34.8 & +38.8{\tiny$\pm$0.6} \\
    \speclevel{2} & 66.0 & 66.8{\tiny$\pm$0.6} & 36.5 & 31.8{\tiny$\pm$0.3} & +29.5 & +35.0{\tiny$\pm$0.6} \\
    \speclevel{3} & 55.8 & 55.2{\tiny$\pm$0.7} & 24.6 & 25.4{\tiny$\pm$0.8} & +31.3 & +29.8{\tiny$\pm$0.3} \\
    \bottomrule
  \end{tabular}
\end{table}

Three patterns emerge. \emph{Single-agent performance is nearly identical}: Sonnet and Haiku trace the same curve from $\sim$88\% at \speclevel{0} to $\sim$55\% at \speclevel{3} (within 2pp at every level), and Haiku's run-to-run standard deviation stays $\leq$0.7pp.

\emph{Split-agent degradation follows the same shape}, though Haiku is slightly more vulnerable at intermediate levels (\speclevel{1}: $36.7 \pm 0.7$\% vs.\ 43.0\%). Across three Haiku runs, split-rate standard deviations remain below 2pp at every level.

\emph{The coordination gap is persistent across models}: Sonnet averages 31.5pp (range 29.5--34.8pp), Haiku averages $32.3 \pm 0.8$pp (range 25.5--38.8pp). While the gap varies more across levels in Haiku, it is always significant and positive. The relevant observation is therefore not that one Claude model outperforms another, but that the same qualitative effect reappears under a different capability profile and across repeated runs. Within the bounds of our design, this supports interpreting the specification gap as a systematic coordination problem rather than an artifact of model capability or sampling noise.

\subsection{Practical Implications}

\begin{enumerate}
  \item \textbf{Specification requirements}: Code agent systems should include data-structure constraints in specifications, not just behavioral descriptions. The recovery experiment confirms that this is sufficient: a merger agent with the full specification recovers single-agent performance.
  \item \textbf{Diagnostic, not therapeutic, conflict detection}: AST-based conflict detection is cheap (zero LLM cost) and valuable for \emph{explaining} failures, but in our experiments it did not enable better merges. Systems should use it as a monitoring signal---not as input to the merger.
  \item \textbf{Architecture guidance}: When specifications are incomplete (\speclevel{2}/\speclevel{3}), the most effective intervention is to enrich the specification, not to add post-hoc conflict analysis. When enrichment is infeasible, consider single-agent generation.
\end{enumerate}

\subsection{Threats to Validity}

\begin{description}
  \item[Internal --- bias injection] Agent biases are injected via system prompt rather than emerging organically. As discussed in \Cref{sec:discussion}, this models a realistic operational risk and the init-visibility experiment confirms that the coordination penalty persists even when both agents share identical constructor information, suggesting the gap is not an artifact of the injection mechanism.

  \item[Internal --- information asymmetry] Single agents receive the \texttt{\_\_init\_\_} body while split agents do not. The init-visibility experiment (\Cref{sec:init-visibility}) disentangles these two factors: coordination accounts for the larger share of the gap (+15.7--19.8pp) while information asymmetry contributes +11.2--15.3pp, with approximately additive interaction ($-4.1$pp).
  \item[Construct] Our central construct is \emph{coordination failure under partial knowledge}. We operationalize it with three complementary signals: integration pass rate, the single-vs.-split gap, and AST-detected structural conflicts. None is a perfect measure on its own. Test pass rate is only a proxy for integration quality: some tests may pass despite latent incompatibilities, or fail for reasons unrelated to coordination. The AST detector's 62.9\% overall recall further shows that many failures are semantic rather than structural. We therefore avoid claiming that every failed merge is caused by the exact conflict types we detect; instead, we interpret convergence across these measures as evidence for a broader coordination construct.
  \item[External] We test two models from the same family (Claude Sonnet 4 and Haiku 4.5) on Python class generation with prompt-induced structural biases. Three independent Haiku runs confirm low aggregate variance ($\leq$2pp), though individual task outcomes can vary substantially (per-task split-rate std up to 51pp). We therefore claim external validity at the level of \emph{mechanism} rather than exact effect size: incomplete specifications reliably expose a coordination penalty in our setting, but the absolute magnitude may change for other model families (\eg, GPT, Gemini), programming languages, orchestration schemes, or less stylized sources of bias. Generalization to those settings requires further study.
\end{description}

% ============================================================
\section{Conclusion}
\label{sec:conclusion}

We studied how specification completeness affects integration success when multiple LLM agents independently generate parts of the same class. Across 51 tasks, two models, and four specification levels, a persistent coordination gap (25--39pp) separates multi-agent from single-agent performance---even when specifications are detailed.

Our experiments reveal an asymmetry between diagnosis and repair. An AST-based conflict detector reliably identifies structural mismatches, and its precision improves precisely where specifications are weakest. Yet providing conflict reports to a merger agent adds no measurable benefit: the full specification alone is sufficient to recover the single-agent ceiling. When both are combined, the redundant diagnostic detail slightly hurts performance. A decomposition experiment further shows that the gap reflects genuine coordination difficulty (+16pp), not only hidden information (+11pp), and that the two effects are approximately additive.

These findings support a \emph{specification-first} view of multi-agent orchestration: the specification functions as a coordination contract between agents, and investing in richer specifications appears more consequential than investing in post-hoc conflict detection. Methodologically, our layered design---specification ablation, factorial recovery, and gap decomposition---provides a reusable template for evaluating coordination in multi-agent code generation. Whether richer forms of conflict representation, such as semantic diffs or iterative negotiation, can partially substitute for specification quality remains an open direction for future work.

% TODO: Add data availability statement and replication package URL

% ============================================================
\bibliographystyle{unsrtnat}
\bibliography{references}

\appendix
\clearpage
\section*{Appendices}
\noindent Additional materials are organized as follows: \Cref{app:taxonomy} provides the full conflict taxonomy; \Cref{app:example} presents a worked example; \Cref{app:spec-levels,app:prompts,app:outputs,app:detection,app:recovery} detail the specification variants, prompts, outputs, conflict reports, and recovery conditions for that example; and \Cref{app:init-visibility} reports the prompts for the init-visibility experiment.

\section{Conflict Taxonomy}
\label{app:taxonomy}

\begin{figure}[h]
  \centering
  \begin{tikzpicture}
    \begin{axis}[
      width=9cm, height=6.5cm,
      ybar stacked,
      bar width=0.7cm,
      axis lines=left,
      xlabel={Specification Level},
      ylabel={Number of Conflicts},
      xtick={0,1,2,3},
      xticklabels={\speclevel{0},\speclevel{1},\speclevel{2},\speclevel{3}},
      ymin=0, ymax=100,
      ytick={0,20,40,60,80,100},
      ymajorgrids=true,
      grid style={dashed, black!20},
      legend style={at={(0.5,-0.28)}, anchor=north, font=\small,
                    draw=none, legend columns=4},
      enlarge x limits=0.3,
      clip=false,
    ]
      % Bottom to top: Type, State, Protocol, Return
      \addplot[fill=colType!70, draw=colType]
        coordinates {(0,26) (1,45) (2,48) (3,53)};
      \addplot[fill=colState!70, draw=colState]
        coordinates {(0,22) (1,23) (2,26) (3,28)};
      \addplot[fill=colProtocol!70, draw=colProtocol]
        coordinates {(0,3) (1,3) (2,3) (3,1)};
      \addplot[fill=colReturn!70, draw=colReturn]
        coordinates {(0,1) (1,1) (2,1) (3,0)};

      % Total labels above stacked bars
      \node[above, font=\scriptsize\bfseries] at (axis cs:0,52) {52};
      \node[above, font=\scriptsize\bfseries] at (axis cs:1,72) {72};
      \node[above, font=\scriptsize\bfseries] at (axis cs:2,78) {78};
      \node[above, font=\scriptsize\bfseries] at (axis cs:3,82) {82};

      \legend{Type, State, Protocol, Return}
    \end{axis}
  \end{tikzpicture}
  \caption{Conflict taxonomy by specification level. Type conflicts (incompatible data structures) dominate at all levels and grow fastest as specifications degrade. Total conflicts increase from 52 at \speclevel{0} to 82 at \speclevel{3}.}
  \label{fig:conflicts}
\end{figure}
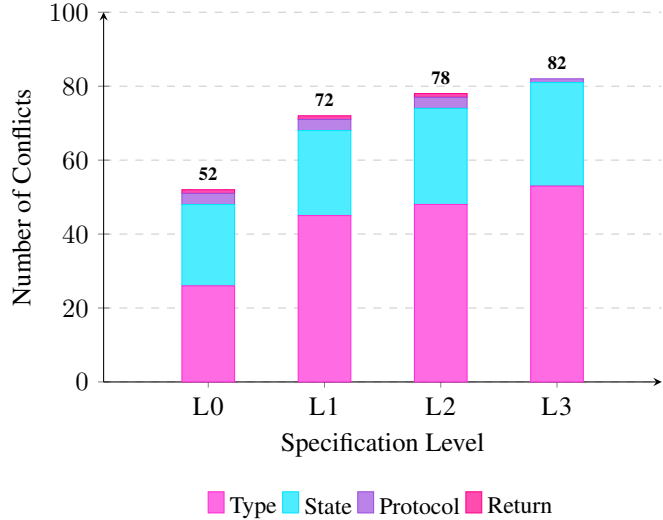

\section{Worked Example: Prompts, Outputs, and Conflicts}
\label{app:example}

This appendix traces a single task---\texttt{AssessmentSystem} (ClassEval\_4, 6 methods, 31 tests)---through every experimental condition, using the exact prompts and real LLM outputs from our experiment. This task exemplifies the specification gap: the single agent passes 97\% of tests, na\"ive merge passes 0\%, and providing the full specification alone recovers to 97\%.

% ── B.1  Specification Levels ──────────────────────────────
\subsection{Specification Levels}
\label{app:spec-levels}

The same class skeleton is presented at two extremes. At \speclevel{0}, the docstrings name the \texttt{students} dict, show its structure via doctests, and describe return types. At \speclevel{3}, only method signatures remain---agents must infer data structures from names alone.

\begin{specbox}[\speclevel{0} Skeleton --- Full Specification]{colNeonPink}
\begin{verbatim}
class AssessmentSystem:
    def __init__(self):
        """Initialize the students dict."""
        self.students = {}

    def add_student(self, name, grade, major):
        """
        Add a new student into self.students dict
        :param name: str, student name
        :param grade: int, student grade
        :param major: str, student major
        >>> system.add_student('student 1', 3, 'SE')
        >>> system.students
        {'student 1': {'name': 'student 1',
          'grade': 3, 'major': 'SE', 'courses': {}}}
        """

    def add_course_score(self, name, course, score):
        """
        Add score of specific course for student
        in self.students
        >>> system.add_course_score('student 1', 'math', 94)
        >>> system.students
        {'student 1': {'name': 'student 1', 'grade': 3,
          'major': 'SE', 'courses': {'math': 94}}}
        """

    def get_gpa(self, name):
        """Get average grade of one student.
        :return: float or None"""

    def get_all_students_with_fail_course(self):
        """Get all students who have any score below 60
        :return: list of str, student name"""

    def get_course_average(self, course):
        """Get average score of a specific course.
        :return: float or None"""

    def get_top_student(self):
        """Find student with highest GPA.
        :return: str, name of student"""
\end{verbatim}
Docstrings explicitly reference ``\texttt{self.students} dict'' and show its nested structure via doctests---the key coordination signal.
\end{specbox}

\begin{specbox}[\speclevel{3} Skeleton --- Bare Signatures]{colElectricBlue}
\begin{verbatim}
class AssessmentSystem:

    def __init__(self):
        self.students = {}

    def add_student(self, name, grade, major):

    def add_course_score(self, name, course, score):

    def get_gpa(self, name):

    def get_all_students_with_fail_course(self):

    def get_course_average(self, course):

    def get_top_student(self):
\end{verbatim}
No docstrings, no type hints. The \texttt{\_\_init\_\_} body hints at a dict, but \textbf{split agents never see it}---it is replaced with \texttt{pass}.
\end{specbox}

% ── B.2  Agent System Prompts ──────────────────────────────
\subsection{Agent System Prompts}
\label{app:prompts}

Each agent receives a short system prompt establishing its role and bias, followed by a user message containing the skeleton and method assignments.

\begin{promptbox}[System Prompt --- Agent A (List Bias)]
You are an expert Python programmer implementing
part of a Python class. Another developer implements
the remaining methods.

Design convention: store collections as LISTS
(sequential, ordered). Use list of dicts for
structured data. Iterate/filter to find entries.

IMPORTANT:
- Define \_\_init\_\_ yourself with appropriate
  instance variables
- Implement ONLY your assigned methods
- For other methods write: def method\_name(self): pass
- Return ONLY Python code, no explanation
\end{promptbox}

\begin{promptbox}[System Prompt --- Agent B (Dict Bias)]
You are an expert Python programmer implementing
part of a Python class. Another developer implements
the remaining methods.

Design convention: store collections as DICTIONARIES
(key-value mapping). Map identifiers to data records
for fast lookup and membership checks.

IMPORTANT:
- Define \_\_init\_\_ yourself with appropriate
  instance variables
- Implement ONLY your assigned methods
- For other methods write: def method\_name(self): pass
- Return ONLY Python code, no explanation
\end{promptbox}

\noindent Split agents receive the \speclevel{3} skeleton \emph{with \texttt{\_\_init\_\_} replaced by \texttt{pass}}. They must define their own constructor. The user message also lists which methods to implement and which belong to the collaborator.

% ── B.3  Agent Outputs ─────────────────────────────────────
\subsection{Agent Outputs (Real LLM Responses)}
\label{app:outputs}

The following are verbatim outputs from Claude Sonnet 4 for this task. Agent~A initializes all fields as \textbf{lists}; Agent~B uses \textbf{dicts}. The conflict is visible in the \texttt{\_\_init\_\_} of each output.

\begin{codebox}[Agent A Output --- List Bias]{colNeonPink}
\begin{verbatim}
class Unknown:
    def __init__(self):
        self.students = []          # <-- LIST
        self.courses = []           # <-- LIST
        self.scores = []            # <-- LIST

    def add_student(self, student_id, name):
        student = {'student_id': student_id,
                   'name': name}
        self.students.append(student)  # list.append()

    def get_gpa(self, student_id):
        student_scores = [s for s in self.scores
                          if s['student_id'] == student_id]
        if not student_scores:
            return 0.0
        return sum(s['score'] for s in student_scores) \
               / len(student_scores)

    def get_course_average(self, course_id):
        course_scores = [s for s in self.scores
                         if s['course_id'] == course_id]
        if not course_scores:
            return 0.0
        return sum(s['score'] for s in course_scores) \
               / len(course_scores)

    # Stubs for collaborator's methods
    def add_course_score(self): pass
    def get_all_students_with_fail_course(self): pass
    def get_top_student(self): pass
\end{verbatim}
\end{codebox}

\begin{codebox}[Agent B Output --- Dict Bias]{colElectricBlue}
\begin{verbatim}
class Unknown:
    def __init__(self):
        self.students = {}          # <-- DICT
        self.courses = {}           # <-- DICT
        self.scores = {}            # <-- DICT (tuple keys)

    def add_course_score(self, student_id,
                         course_id, score):
        self.scores[(student_id, course_id)] = score

    def get_all_students_with_fail_course(self,
            failing_threshold=60):
        failing = {}
        for (sid, cid), score in self.scores.items():
            if score < failing_threshold:
                if sid not in failing:
                    failing[sid] = []
                failing[sid].append(cid)
        return failing

    def get_top_student(self):
        if not self.scores:
            return None
        totals, counts = {}, {}
        for (sid, cid), score in self.scores.items():
            totals[sid] = totals.get(sid, 0) + score
            counts[sid] = counts.get(sid, 0) + 1
        best, best_avg = None, -1
        for sid in totals:
            avg = totals[sid] / counts[sid]
            if avg > best_avg:
                best_avg, best = avg, sid
        return best

    # Stubs for collaborator's methods
    def add_student(self): pass
    def get_gpa(self): pass
    def get_course_average(self): pass
\end{verbatim}
\end{codebox}

\noindent\textbf{Na\"ive merge result: 0/31 tests pass.} When Agent~A's list-based \texttt{\_\_init\_\_} is combined with Agent~B's dict-based methods, every operation crashes: \texttt{self.scores[(sid, cid)] = score} fails on a list, and \texttt{self.students.append()} fails on a dict.

% ── B.4  AST Conflict Detection ────────────────────────────
\subsection{AST Conflict Detection Output}
\label{app:detection}

The AST detector compares the two agents' code \emph{before integration} and reports:

\begin{conflictbox}[AST Conflict Report --- 4 Conflicts Detected (3 Type + 1 State)]
\begin{verbatim}
Conflict 1 [TYPE, HIGH]: field self.students
  Agent A: Initializes as list
  Agent B: Initializes as dict

Conflict 2 [TYPE, HIGH]: field self.courses
  Agent A: Initializes as list
  Agent B: Initializes as dict

Conflict 3 [TYPE, HIGH]: field self.scores
  Agent A: Initializes as list
  Agent B: Initializes as dict

Conflict 4 [STATE, LOW]: field self.students
  Cross-boundary state dependency with mutations
  Agent A: Operations: ['append']
  Agent B: Operations: read/write only
\end{verbatim}
All three type conflicts are correctly identified. The state conflict detects that Agent~A mutates \texttt{self.students} with \texttt{append} (a list operation), while Agent~B only reads/writes it as a dict.
\end{conflictbox}

% ── B.5  Recovery Conditions ───────────────────────────────
\subsection{Recovery Conditions}
\label{app:recovery}

The merger agent receives the two conflicting implementations and must produce a unified class. The key difference between conditions is what information the merger receives.

\begin{promptbox}[Spec-Only Merger --- \speclevel{0} Spec, No Conflict Report]
You are an expert Python programmer. Combine two
developers' implementations into a single correct
class. Use the full specification to guide your
merge. Return ONLY Python code.

\textcolor{colNeonPink!70!black}{--- User message includes L0 skeleton (with docstrings)}
\textcolor{colNeonPink!70!black}{--- + Agent A code + Agent B code}
\textcolor{colNeonPink!70!black}{--- NO conflict report provided}
\end{promptbox}
\centerline{\textbf{Result: 96.8\% (30/31 tests) --- full recovery.}}

\medskip

\begin{promptbox}[Guided Merger --- \speclevel{3} Spec, With Conflict Report]
You are an expert Python programmer. Combine two
developers' implementations into a single correct
class. Pay attention to the detected conflicts.
Return ONLY Python code.

\textcolor{colElectricBlue!60!black}{--- User message includes L3 skeleton (bare signatures)}
\textcolor{colElectricBlue!60!black}{--- + Agent A code + Agent B code}
\textcolor{colElectricBlue!60!black}{--- + full conflict report (see B.4)}
\end{promptbox}
\centerline{\textbf{Result: 6.5\% (2/31 tests) --- conflict report hurts without spec.}}

\bigskip

\Cref{tab:example-results} summarizes the results for this task across all conditions.

\begin{table}[h]
  \caption{Results for AssessmentSystem across all experimental conditions (31 tests).}
  \label{tab:example-results}
  \centering
  \begin{tabular}{llcc}
    \toprule
    Condition & Spec / Conflicts & Tests & Pass rate \\
    \midrule
    Single     & \speclevel{0} / ---    & 30/31 & 96.8\% \\
    Na\"ive    & (concatenation)        & 0/31  & 0.0\% \\
    \midrule
    Blind      & \speclevel{3} / no     & 14/31 & 45.2\% \\
    Guided     & \speclevel{3} / yes    & 2/31  & 6.5\% \\
    Spec-Only  & \speclevel{0} / no     & 30/31 & 96.8\% \\
    Resolve    & \speclevel{0} / yes    & 30/31 & 96.8\% \\
    \bottomrule
  \end{tabular}
\end{table}

The pattern is stark: the \speclevel{0} specification alone enables full recovery (Spec-Only = Single = 96.8\%), while the conflict report without the specification actually \emph{hurts} (Guided: 6.5\% vs.\ Blind: 45.2\%). With the full specification available, the merger infers the correct dict-based structure from the docstring ``Add a new student into \texttt{self.students} dict'' without needing any explicit conflict report.

\section{Init Visibility Experiment Prompts}
\label{app:init-visibility}

This appendix shows the prompts used in the init-visibility experiment (\Cref{sec:init-visibility}).  The key difference from the main experiment is that the ``visible'' conditions provide agents with the real \texttt{\_\_init\_\_} body.

\subsection{Single-Hidden Condition}

The single agent receives the skeleton with \texttt{\_\_init\_\_} replaced by \texttt{pass}---identical to what split agents see in the main experiment---and must define its own data structures without any bias.

\begin{promptbox}[System Prompt --- Single Agent (Hidden Init)]
You are an expert Python programmer.
Implement all methods of the given class.
Define \_\_init\_\_ yourself with appropriate
instance variables.
Return ONLY Python code, no explanation.
\end{promptbox}

\subsection{Split-Visible Condition}

Split agents receive the \textit{full} skeleton including the \texttt{\_\_init\_\_} body, and are explicitly told to keep it exactly as provided.  Crucially, they still carry the list/dict bias---the question is whether the visible constructor overrides their prior.

\begin{promptbox}[System Prompt --- Agent A (Visible Init, List Bias)]
You are an expert Python programmer implementing
part of a Python class. Another developer
implements the remaining methods.

Design convention: store collections as LISTS
(sequential, ordered). Use list of dicts for
structured data. Iterate/filter to find entries.

IMPORTANT:
- Keep \_\_init\_\_ EXACTLY as provided
- Implement ONLY your assigned methods
- For other methods write: def method\_name(self): pass
- Return ONLY Python code, no explanation
\end{promptbox}

\noindent The user message for split-visible agents includes the constructor body explicitly:

\begin{specbox}[User Message Excerpt --- Split-Visible Agent]{colElectricBlue}
\begin{verbatim}
Class: AssessmentSystem
Description: ...

Constructor (keep EXACTLY as provided):
    def __init__(self):
        self.students = {}

YOUR methods to implement:
    def add_student(self, name, grade, major):
        """Add a new student..."""
    ...
\end{verbatim}
\end{specbox}

\noindent By providing the real \texttt{\_\_init\_\_} body, this condition tests whether a shared constructor can anchor agents to a common data structure despite their opposing biases.

\end{document}